# Metallic ground state in an ion-gated two-dimensional superconductor

Yu Saito,[1] Yuichi Kasahara,[1,2]* Jianting Ye,[1,3,4]* Yoshihiro Iwasa,[1,4]† Tsutomu Nojima[5]†

[1] Quantum-Phase Electronics Center (QPEC) and Department of Applied Physics, The University of Tokyo, Tokyo 113-8656, Japan
[2] Department of Physics, Kyoto University, Kyoto 606-8502, Japan
[3] Zernike Institute for Advanced Materials, University of Groningen, Nijenborgh 4, 9747 AG Groningen, The Netherlands
[4] RIKEN Center for Emergent Matter Science (CEMS), Wako 351-0198, Japan
[5] Institute for Materials Research, Tohoku University, Sendai 980-8577, Japan

*These authors contributed equally to this work.
†Corresponding author. E-mail: iwasa@ap.t.u-tokyo.ac.jp (Y.I.); nojima@imr.tohoku.ac.jp (T.N.)

**Recently emerging two-dimensional (2D) superconductors in atomically thin layers and at heterogeneous interfaces are attracting growing interest in condensed matter physics. Here, we report that an ion-gated ZrNCl surface, exhibiting a dome-shaped phase diagram with a maximum critical temperature of 14.8 kelvin, behaves as a superconductor persisting to the 2D limit. The superconducting thickness estimated from the upper critical fields is $\cong$ 1.8 nanometers, which is thinner than one-unit-cell. The majority of the vortex phase diagram down to 2 kelvin is occupied by a metallic state with a finite resistance, owing to the quantum creep of vortices caused by extremely weak pinning and disorder. Our findings highlight the potential of electric-field-induced superconductivity, establishing a new platform for accessing quantum phases in clean 2D superconductors.**

Recent technological advances of materials fabrication have led to discoveries of a variety of superconductors at heterogeneous interfaces and in ultrathin films; examples include superconductivity at oxide interfaces (*1*, *2*), electric-double-layer interfaces (*3*), and mechanically cleaved (*4*), molecular-beam-epitaxy grown (*5*,*6*), or chemical vapor deposited (*7*) atomically thin layers. These systems are providing opportunities for searching for superconductivity at higher temperatures as well as investigating the intrinsic nature of two-dimensional (2D) superconductors, which are distinct from bulk superconductors because of enhanced thermal and quantum fluctuations.

One of the issues to be addressed is how zero electrical resistance is achieved or destroyed. In 2D superconductors exposed to magnetic fields, vortex pinning, which is necessary to achieve zero electrical resistance under magnetic fields, may be too weak. To address this question, superconductivity in vacuum-deposited metallic thin films has been studied (*8*). The conventional way to approach the 2D limit is to reduce the film thickness, but this concomitantly increases disorder. In such thin films, a direct superconductor-to-insulator transition (SIT) is observed in most cases when the system is disordered (*8*).This SIT has been understood in the framework of the so-called "dirty boson model" (*9*); however, this simple picture needed to be modified because an intervening metallic phase between the superconducting (SC) and insulating phases under magnetic fields was observed in less-disordered systems (*8, 10*). Therefore, investigating 2D superconductivity in even cleaner systems is desirable.

In contrast to the conventional metallic films, the recently discovered 2D superconductors are highly crystalline, displaying lower normal state sheet resistance. The electric-double-layer transistor (EDLT), which is composed of the interface between crystalline solids and electrolytes, can be a good candidate to realize such a clean system, since the conduction carriers are induced electrostatically at the atomically flat surfaces without introducing extrinsic disorder. In addition, the EDLT has the greatest advantage of their applicability to a wide range of materials, as exemplified by gate-induced superconductivity in three-dimensional $SrTiO_3$ (*3*, *11*, *12*), $KTaO_3$ (*13*), quasi-2D layered ZrNCl (*14*), transition metal dichalcogenides (*15-17*) and cuprates (*18-22*). Here, we report comprehensive transport studies on a ZrNCl-EDLT, which provide evidence of 2D superconductivity in this system based on several types of analyses. In particular, we found that the zero resistance state is immediately destroyed by the application of finite out-of-plane magnetic fields, and consequently, a metallic state is stabilized in a wide range of magnetic fields. This is a manifestation of the quantum tunneling of vortices due to the extremely weak pinning in the ultimate 2D system.

ZrNCl is originally an archetypal band insulator with a layered crystal structure (*23, 24*), in which a unit cell comprises three $(ZrNCl)_2$ layers (Fig. 1A). Bulk ZrNCl becomes a superconductor with a critical temperature, $T_c$, as high as 15.2 K by alkali–metal intercalation (*25-28*). Figure 1B shows the relation between the sheet conductance, $\sigma_{sheet}$, and the gate voltage, $V_G$, for a ZrNCl-EDLT with a 20-nm-thick flake without any monolayer steps (*29*), measured at a source–drain voltage of $V_{DS} = 0.1$ V and at a temperature of $T = 220$ K. $\sigma_{sheet}$ abruptly increased at $V_G > 2$ V, demonstrating a typical *n*-type field-effect transistor behavior. As shown in the temperature dependence of the sheet resistance, $R_{sheet}$, at different $V_G$ values (Fig. 1C), the insulating phase is dramatically suppressed with increasing $V_G$, and finally a resistance drop due to a SC transition appears at $V_G = 4$ V. Zero resistance (below ~ 0.05 Ω) was achieved at $V_G = 6$ and 6.5 V. Despite such relatively large gate voltages, any signature of electrochemical process was not observed (figs. S1, S2) (*29*). The tail of the resistance drop at 6.5 V can be explained in terms of the Berezinskii–Kosterlitz–Thouless (BKT) transition (Fig. 1D), which realizes a zero-ohmic-resistance phase driven by the binding of vortex-antivortex pairs. To determine the BKT transition temperature, we use the Halperin–Nelson equation (*30*, *31*), which shows a square-root-cusp behavior that originates from the energy dissipation due to the Bardeen-Stephen vortex flow above the BKT transition temperature. On the other hand, a gradual decrease of $R_{\text{sheet}}$ at temperatures far above $T_c$, leading to a broadened SC onset (Fig. 1D, inset), was observed. This feature can be well reproduced by an analysis that takes both the Aslamazov–Larkin and Maki–Thompson terms (*32-34*) for the 2D fluctuation conductivities into account (fig. S4) (*29*). These results suggest that 2D superconductivity is achieved at zero magnetic field.

Figures 2A and B display temperature-dependent $R_{\text{sheet}}(T)$ values at $V_G = 6.5$ V for magnetic fields applied perpendicular and parallel to the surface of ZrNCl, respectively. For the out-of-plane magnetic fields, $T_c$ is dramatically suppressed with a considerable broadening of the SC transition even at a small magnetic field of $\mu_0 H = 0.05$ T, which is in marked contrast to those observed in the in-plane magnetic field geometry (see also fig. S5). Such a large anisotropy suggests that the superconductivity is strongly 2D in nature and indicates a significant contribution of the vortex motion in the out-of-plane magnetic field geometry. Figure 2C shows the angular dependence of the upper critical field, $\mu_0 H_{c2}(\theta)$ ($\theta$ represents the angle between the c-axis of ZrNCl and applied magnetic field directions), at 13.8 K, which is just below $T_c$ (= 14.5 K at zero magnetic field). $T_c$ ($H_{c2}$) was defined as the temperature (magnetic field) where $R_{\text{sheet}}$ becomes 50 % of the normal state resistance, $R_N$, at 30 K (*29*). A cusp-like peak is clearly resolved at $\theta = 90°$ (Fig. 2C, inset), and is qualitatively distinct from the three-dimensional anisotropic mass model but is well described by the 2D Tinkham model (*35*). Similar observations have been reported in a SrTiO$_3$-EDLT (*36*), implying that the EDLT is a versatile tool for creating 2D superconductors. Figure 2D shows the temperature dependence of $\mu_0 H_{c2}$ at $\theta$

= 90° ($\mu_0 H_{c2}^{\parallel}$) and at $\theta = 0°$ ($\mu_0 H_{c2}^{\perp}$), which exhibits a good agreement with the phenomenological Ginzburg-Landau (GL) expressions for 2D SC films:

$$\mu_0 H_{c2}^{\perp} = \frac{\Phi_0}{2\pi \xi_{GL}(0)^2}\left(1 - \frac{T}{T_c}\right) \tag{1}$$

$$\mu_0 H_{c2}^{\parallel} = \frac{\Phi_0 \sqrt{12}}{2\pi \xi_{GL}(0) d_{SC}} \sqrt{1 - \frac{T}{T_c}} \tag{2}$$

where $\Phi_0$ is the flux quantum, $\xi_{GL}(0)$ is the extrapolation of the GL coherence length, $\xi_{GL}$, at $T = 0$ K, and $d_{sc}$ is the temperature-independent SC thickness. As a result of the fit, we obtained $\xi_{GL}(0) \cong 12.8$ nm and $d_{sc} \cong 1.8$ nm. The latter parameter approximately corresponds to the bilayer thickness of the (ZrNCl)$_2$ layer, which is less than one-unit-cell thick. The estimated thickness is indeed in the atomic scale, and demonstrates that the superconductivity persists to the extreme 2D limit. The $d_{sc}$ for this system is much smaller than the reported value of ~11 nm for the interface superconductivity on SrTiO$_3$ (*1, 36*), which is presumably owing to the huge dielectric constant in the incipient ferroelectric SrTiO$_3$. Recently, it was suggested based on theoretical calculations that the depth of the induced charge carriers in ion-gated superconducting ZrNCl is only one layer (*37*). The difference from the present observation might be ascribed to the proximity effect of the superconductivity, which is a phenomenon whereby the Cooper pairs in a SC layer (the topmost layer, in the present case) diffuse into the neighboring non-SC layers (the second layer), resulting in broadening of the effective thickness. This could occur even if there are only a small number of electrons in the second layer. Another possibility for this discrepancy might come from the situation that the measured $H_{c2}$ is suppressed because of the paramagnetic effect as compared with the orbital limit, leading to an estimated $d_{sc}$ thicker than the real value.

In the present system, we found that the Pippard coherence length, $\xi_{\text{Pippard}}$, is equal to 43.4 nm, as calculated from $\xi_{\text{Pippard}} = \hbar v_F / \pi \Delta(0)$ by using $v_F = \hbar k_F / m^*$, $k_F = (4\pi n_{2D}/ss')^{1/2}$ and the Bardeen–Cooper–Schrieffer (BCS) energy gap of $\Delta(0) = 1.76 k_B T_c = 2.2$ meV, where $v_F$, $k_F$, $m^*$, $\hbar$, $s$ and $s'$ are the Fermi velocity, the Fermi wave number, the effective mass, Planck's constant divided by $2\pi$, the spin degree of freedom and the valley degree of freedom, respectively, for $V_G = 6.5$ V (the sheet carrier density of $n_{2D} = 4.0 \times 10^{14}$ cm$^{-2}$) with $T_c = 14.5$ K and the effective mass of $m^* = 0.9 m_0$ (38). Here $s$ and $s'$ are both 2. The Pippard coherence length is larger than $k_F^{-1} = 0.28$ nm, and much larger than $d_{\text{sc}} \cong 1.8$ nm. We also note that the $H_{c2}^{\parallel}$ may exceed the Pauli limit for weak coupling BCS superconductors, $\mu_0 H_P^{\text{BCS}} = 1.86 T_c = 27.0$ T. However, to confirm this phenomenon, it is necessary to investigate $H_{c2}$ at lower temperatures by using high magnetic fields.

Having estimated $d_{\text{sc}}$, we can now compare the phase diagrams of electric-field-induced 2D and bulk superconductors (28) (Fig. 3). A direct comparison is made by using $n_{2D}$ estimated from Hall-effect measurements in ZrNCl-EDLTs (fig. S6) (29) and $n_{2D}$ in the (ZrNCl)$_2$ bilayer for the bulk. In contrast to the bulk, where $T_c$ abruptly appears at $n_{2D} = 1.5 \times 10^{14}$ cm$^{-2}$ followed by a decrease with increasing $n_{2D}$, ZrNCl-EDLTs exhibit a gradual increase of $T_c$, forming a dome-like SC phase with a maximum $T_c$ of 14.8 K at $n_{2D} = 5.0 \times 10^{14}$ cm$^{-2}$. The different phase diagrams between electric-field-induced and intercalated superconductivity in Fig. 3 suggest the importance of two dimensionality and broken inversion symmetry in the presence of an electric field, which may lead to exotic superconducting phenomena such as the spin-parity mixture state and the helical state. On the other hand, the coincidence of the critical $n_{2D}$ in the 2D and bulk system implies that the mysterious phase diagram in the bulk Li$_x$ZrNCl; that is, an abrupt drop of $T_c$ near $n_{2D} \sim 1\times10^{14}$ cm$^{-2}$, might be related to the quantum SIT phenomena realized in the 2D limit (8). Similar dome-like SC phases have already been reported in EDLTs based on the band insulators KTaO$_3$ (13) and MoS$_2$ (15), suggesting a commonality among electric-field-induced superconductors.

In an Arrhenius plot of $R_{sheet}(T)$ for out-of-plane magnetic fields at $V_G = 6.5$ V (Fig. 4A), $R_{sheet}(T)$ exhibits an activated behavior just below $T_c$ described by $R = R'\exp(-U(H)/k_B T)$, where $k_B$ is Boltzmann's constant, as shown by the dashed lines. The magnetic field dependence of the extracted activation energy $U(H)$ (Fig. 4B), and the relation between $U(H)/k_B T_c$ and $\ln R'$ (Fig. 4B, inset) are consistent with the thermally-assisted collective vortex-creep model in two dimensions (39), $U(H) \propto \ln(H_0/H)$ and $\ln R' = U(H)/k_B T_c + \text{const.}$. The activation energy becomes almost zero at $\mu_0 H \cong 1.3$ T, allowing the vortex flow motion above this field as explained below.

At low temperatures, on the other hand, each $R_{sheet}$–$T$ curve clearly deviates from the activated behavior and then is flattened at a finite value down to the lowest temperature (= 2 K) even at $\mu_0 H$ (= 0.05 T) $\sim \mu_0 H_{c2}/40$. This implies that a metallic ground state exists for at least $\mu_0 H > 0.05$ T and may be a consequence of the vortex motion driven by quantum mechanical processes. Our results are markedly distinct from conventional theories that predict a vortex-glass state and a direct SIT at $T = 0$ with $R_N$ close to the quantum resistance $R_Q = h/4e^2 =$ 6.45 k$\Omega$, where $h$ and $e$ are Plank's constant and the elementary charge, respectively. The metallic ground state has been reported in MoGe (10) and Ta (40) thin films, where $R_N$ values ($\geq 1$ k$\Omega$) are smaller than $R_Q$. In the ZrNCl-EDLT, $R_N$ at $V_G = 6$ and 6.5 V is $\sim 120$ and $\sim 200$ $\Omega$, respectively, which are even lower, reaching values as small as $\sim 1/50$ of $R_Q$. This leads to $k_F l =$ $\frac{1}{ss'}\frac{2h}{e^2}\frac{1}{R_N} \sim 77 - 130$ (for $R_N \sim 120 - 200\ \Omega$) with the mean free path $l$, which is much larger than the Ioffe-Regel limit ($k_F l \sim 1$), indicating that the normal states are relatively clean. We also note that the estimated values of $l \sim 35$ nm (6 V) and 18 nm (6.5 V) result in the relation $\xi_{Pippard} \cong 1.1 - 2.4\ l$, which is far from the dirty limit ($\xi_{Pippard} \gg l$). Indeed, the expression for $\xi_{GL}(0)$ in the dirty limit, $0.855\,(\xi_{Pippard}\,l)^{1/2}$, is not applicable to our result. Furthermore, our $R_{sheet}$-$T$ data under magnetic fields do not follow the scaling relations for a magnetic-field-induced SIT, which has been demonstrated in disordered systems (*8*). All of these features indicate that the ZrNCl-EDLT at $V_G = 6.5$ V is out of the disordered regime and may be entering a moderately clean regime with weak pinning.

The most plausible description of the metallic state is temperature-independent quantum tunneling of vortices (quantum creep). In this model, the resistance obeys a general form in the limit of the strong dissipation (*41*):

$$R_{sheet} \sim \frac{\hbar}{4e^2}\frac{\kappa}{1-\kappa},\qquad \kappa \sim \exp\left\{C\frac{\hbar}{e^2}\frac{1}{R_N}\left(\frac{H-H_{c2}}{H}\right)\right\} \qquad (3)$$

where $C$ is a dimensionless constant. As seen in Fig. 4C, the $R_{sheet}-H$ relation at 2 K is well fitted by eq. (3) up to 1.3 T, indicating that the quantum creep model holds for a wide range. It should be noted that (a) finite-size effects (*42*) or, (b) the model of random Josephson junction arrays that originate from surface roughness (for example, amorphous Bi thin films) (*31*) or inhomogeneous carrier accumulation, both of which can cause the flattening of the resistance with a finite value, can be excluded because of following reasons: (a) a BKT transition, and a zero resistance state (below $\sim 0.05\ \Omega$) are observed (see Fig. 1D). (b) The channel surface preserved atomically-flat morphology with the average mean square roughness of $\sim 0.068$ nm, which is less than 4% of $d_{sc}$, even after all the measurements (fig. S3) (*29*). Also, all the different voltage probes used to measure the longitudinal resistances and the tranverse Hall resistances in four-terminal goemetry (*29*) showed almost the same values with the differences of less than 5 % at the high carrier concentrations ($V_G$ = 6 and 6.5 V), which suggests that the surface carrier accumulation in the present system is homogeneous. Above 1.3 T, $R_{sheet}$ is well described by an *H*-linear dependence (Fig. 4C, inset), indicating pinning-free vortex flow. The crossover from the creep to the flow motion occurs at $\cong$ 1.3 T, where $U(H)$ for the thermal creep approaches zero (Fig. 4B), implying that the pinning or the elastic potential effectively disappears at high magnetic fields. Based on the above observations, we obtained the field-temperature phase diagram of ion-gated ZrNCl shown in Fig. 4D. A true zero resistance state occurs only at very small magnetic fields below 0.05 T. The key parameters here are the dimensionality and $R_N$ of the system. The former enhances the quantum fluctuations, whereas the latter controls the coupling of vortices to a dissipative bath, which stabilizes the metallic region when $R_N$ is low (*43*). Our results indicate that the EDLT provides a model platform of clean 2D superconductors with weak pinning and disorder, which may potentially lead to realizing intrinsic quantum states of matter.

**Acknowledgments**

We thank M. Nakano and Y. Nakagawa for technical support, M. Yoshida for fruitful discussions, and N. Shiba for his useful comments on the manuscript. Y.S. was supported by the Japan Society for the Promotion of Science (JSPS) through a research fellowship for young scientists. J.T.Y. acknowledges the funding from the European Research Council (consolidator grant no. 648855 Ig-QPD). This work was supported by the Strategic International Collaborative Research Program (SICORP-LEMSUPER) of the Japan Science and Technology Agency, Grant-in-Aid for Specially Promoted Research (no. 25000003) from JSPS and Grant-in Aid for Scientific Research on Innovative Areas (no. 22103004) from MEXT of Japan.

**Supplementary Materials**

www.sciencemag.org/

Materials and Methods

Supplementary text

Figs. S1 to S6

References (*44–49*)

# Figure Captions

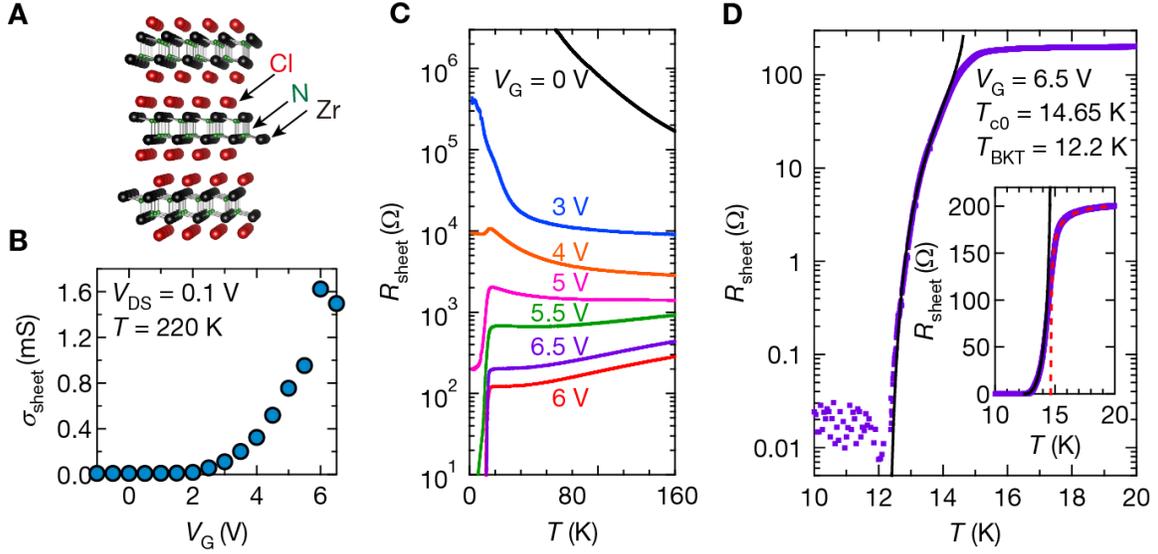

**Fig. 1. Crystal structure of ZrNCl and transport properties of a ZrNCl-EDLT.** (**A**) Ball-and-stick model of a ZrNCl single crystal. The monolayer is 0.92 nm thick. (**B**) Sheet conductance, $\sigma_{sheet}$, of a ZrNCl-EDLT as a function of gate voltage, $V_G$, at 220 K. (**C**) Temperature, $T$, dependence of the sheet resistance, $R_{sheet}$, at different gate voltages, $V_G$, from 0 V to 6.5 V. The device was cooled down to low temperatures after applying $V_G$ gate voltages at 220 K. (**D**) Resistive transition at zero magnetic field and $V_G = 6.5$ V, plotted on a semi-log scale (linear scale in the inset). The black solid line represents the BKT transition using the Halperin–Nelson equation (*30*), $R = R_0 \exp\left\{-2b\left(\dfrac{T_{c0}-T}{T-T_{BKT}}\right)^{1/2}\right\}$, where $R_0$ and $b$ are material parameters. This gives a BKT transition temperature of $T_{BKT} = 12.2$ K with $b = 1.9$. The red dashed line in the inset represents the superconducting amplitude fluctuation taking into account both the 2D Aslamazov–Larkin (*32*) and Maki–Thompson terms (*33*, *34*), which give the temperature, $T_{c0}$, at which the finite amplitude of the order parameter develops (fig. S4) (*29*).

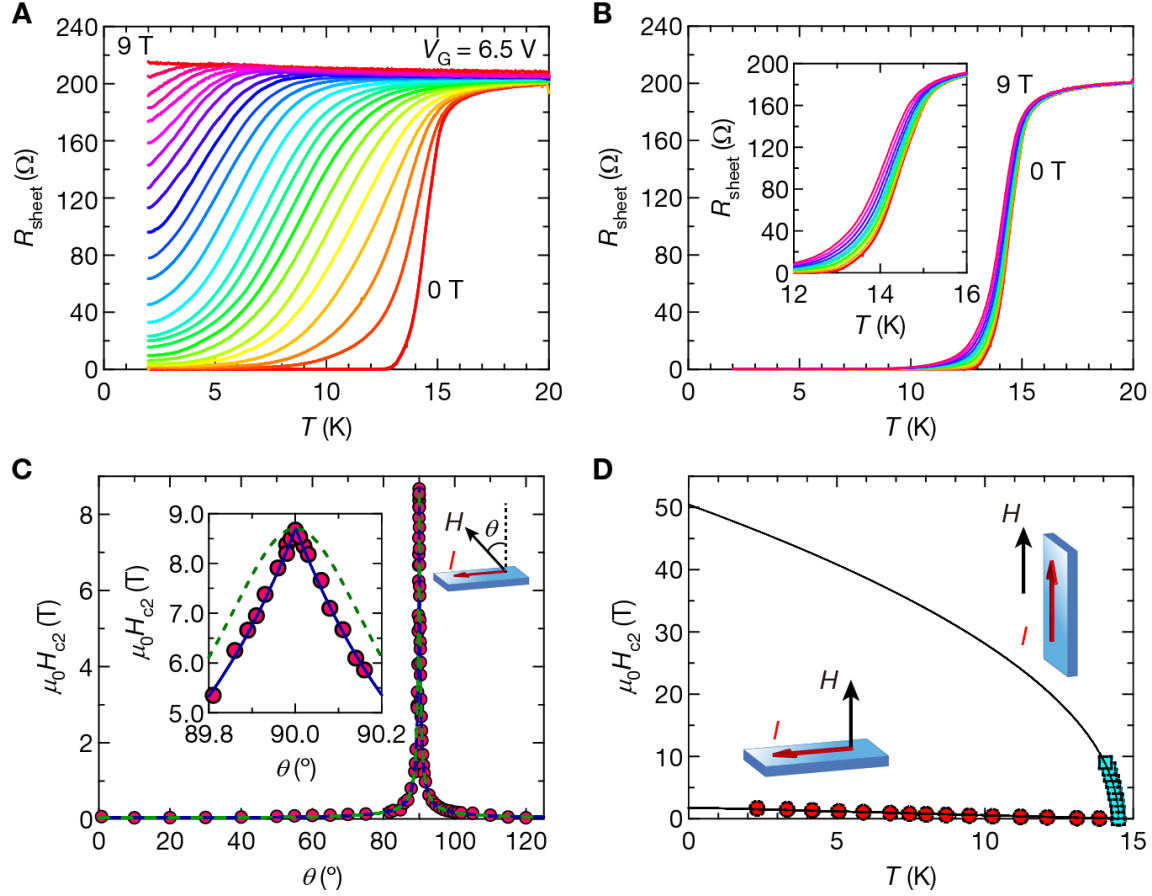

**Fig. 2. Two-dimensional superconductivity in ion-gated ZrNCl.** (**A**) (**B**) Sheet resistance of a ZrNCl-EDLT as a function of temperature at $V_G = 6.5$ V, for (**A**) perpendicular magnetic fields, $\mu_0 H_{c2}^{\perp}$, varying in 0.05 T steps from 0 to 0.1 T, in 0.1 T steps from 0.1 to 0.9 T, and in 0.15 T steps from 0.9 to 2.7 T, and of 3 T and 9 T, and (**B**) parallel magnetic fields, $\mu_0 H_{c2}^{\parallel}$, varying in 1 T steps from 0 to 9 T, respectively. The inset of Fig. 2B is a magnified view of the region between 12 and 16 K. (**C**) Angular dependence of the upper critical fields $\mu_0 H_{c2}(\theta)$ ($\theta$ represents the angle between a magnetic field and the perpendicular direction to the surface of ZrNCl). The inset shows a close-up of the region around $\theta = 90°$. The blue solid line and the green dashed line are the theoretical representations of $H_{c2}(\theta)$, using the 2D Tinkham formula $\left(H_{c2}(\theta)\sin\theta/H_{c2}^{\parallel}\right)^2 + \left|H_{c2}(\theta)\cos\theta/H_{c2}^{\perp}\right| = 1$ and the three-dimensional anisotropic mass model

$H_{c2}(\theta) = H_{c2}^{\parallel} / \left( \sin^2\theta + \gamma^2 \cos^2\theta \right)^{1/2}$ with $\gamma = H_{c2}^{\parallel} / H_{c2}^{\perp}$, respectively. (**D**) Temperature dependence of $\mu_0 H_{c2}$ perpendicular and parallel to the surface, $\mu_0 H_{c2}^{\perp}(T)$ and $\mu_0 H_{c2}^{\parallel}(T)$. Solid black curves are theoretical curves obtained from the 2D Ginzburg–Landau equation.

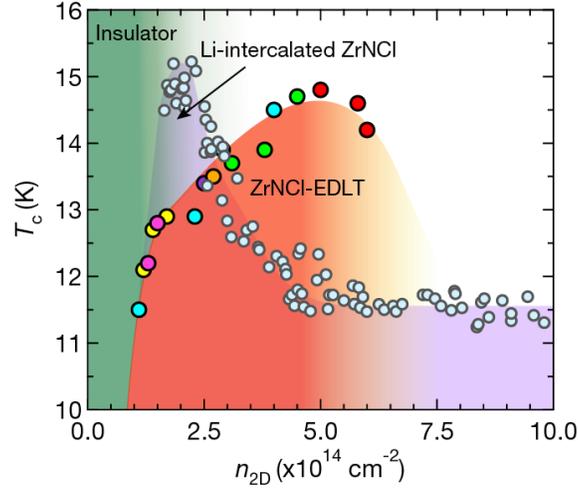

**Fig.3. Electronic phase diagram of electric-field-induced and bulk superconductivity in ZrNCl**. The SC transition temperatures, $T_c$ (defined as the temperature at which $R_{sheet}$ reaches half the value of $R_N$ at 30 K), for ZrNCl-EDLTs were measured in seven different devices, indicated by the colored circles. The data for Li-intercalated ZrNCl was taken from (*28*), and the thickness of a bilayer was used to calculate $n_{2D}$ from the 3D carrier density. The sheet carrier density for ZrNCl-EDLTs was determined by Hall-effect measurements at 60 K (fig. S6) (*29*).

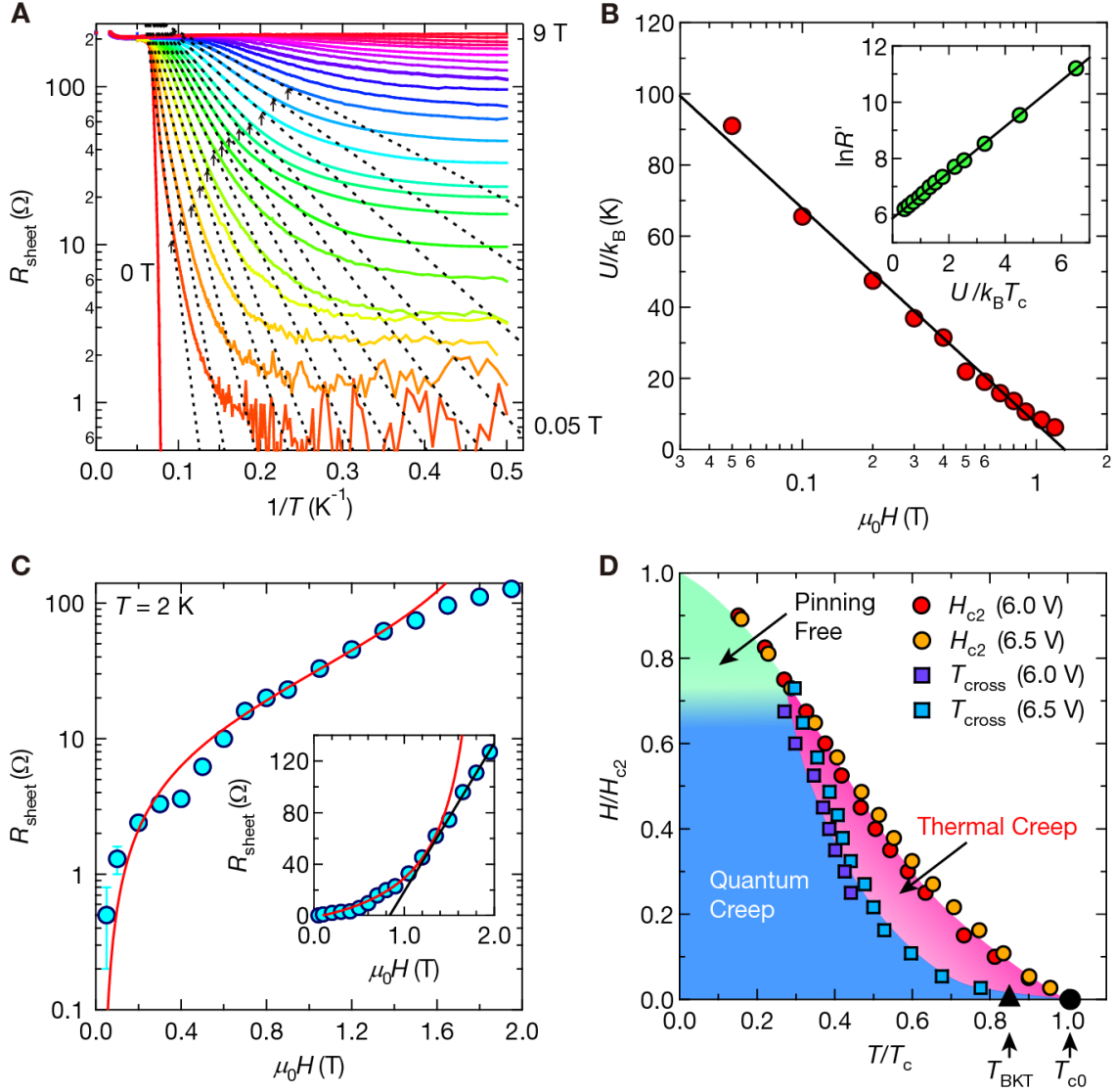

**Fig. 4. Vortex dynamics in ion-gated ZrNCl.** (**A**) Arrhenius plot of the sheet resistance of a ZrNCl-EDLT at $V_G$ = 6.5 V for different magnetic fields perpendicular to the surface of ZrNCl. The black dashed lines demonstrate the activated behavior described by $R_{sheet} = R' \exp(-U(H)/k_B T)$. The arrows separate the thermally activated state in the high-temperature limit and the saturated state at lower temperatures. (**B**) Activation energy, $U(H)/k_B$, which is derived from the slopes of the dashed lines in Fig. 4A, is shown on a semi-logarithmic plot as a function of magnetic field. The solid line is a fit by using the equation,

$U(H) = U_0 \ln(H_0 / H)$. The inset shows the same data plotted as ln$R$' versus $U/k_B T_c$. These plots indicate that the resistance is governed by the thermally activated motion of dislocations of the 2D vortex lattice (2D thermal collective creep). (**C**) Low-temperature saturated values of the resistance as a function of magnetic field at 2 K. The red solid line is a fit using eq. (3), which gives $C \cong 4.8 \times 10^{-3}$. The inset shows $H$-linear dependence of $R_{sheet}$ above 1.3 T (black solid line). (**D**) Vortex phase diagram of the ZrNCl-EDLT. The boundary between the thermally-assisted vortex-creep (thermal creep) regime and the quantum creep regime, $T_{cross}$, is determined from the Arrhenius plot as shown by the arrows in Fig. 4A. $T_{BKT}$ and $T_{c0}$ are the BKT transition temperature and the temperature obtained by analyses of the superconducting amplitude fluctuation, respectively (fig. S4) (*29*).

# Supplementary Materials for

## Metallic ground state in an ion-gated two-dimensional superconductor

Yu Saito, Yuichi Kasahara, Jianting Ye, Yoshihiro Iwasa,* Tsutomu Nojima*

*Corresponding author. E-mail: iwasa@ap.t.u-tokyo.ac.jp (Y.I.); nojima@imr.tohoku.ac.jp (T.N.)

**This PDF file includes:**

    Materials and Methods
    Supplementary Text
    Figs. S1 to S6
    Full Reference List

**Materials and Methods**

Crystal growth and device fabrication

On the whole, we followed the method given in (*14*) for fabricating the devices. Pristine ZrNCl single crystals were grown by a chemical vapor transport method (*25*). We cleaved bulk ZrNCl single crystals into multilayer flakes using the Scotch-tape method, and then transferred those flakes onto a Si/SiO$_2$ substrate. Au (100 nm)/Ti (5 nm) electrodes were patterned onto an isolated thin flake in a Hall bar configuration, and a side gate electrode was patterned on the Si/SiO$_2$ substrate. We covered the device with ZEP 520A (a resist used in electron-beam lithography), except for the EDLT channel surface (Fig. S3A), to avoid electrochemical intercalation from the edges of the flake. A droplet of ionic liquid covered both the channel area and the gate electrode. Ionic liquid N,N-diethyl-N-(2-methoxyethyl)-N- methylammonium bis (trifluoromethylsulphonyl) imide (DEME-TFSI) with the grass transition temperature of ~ 190 K, was selected as a gate dielectric, because it has a large electrochemical window of up to ~ ± 3 V at room temperature. The electrochemical window can be even larger, up to ~ ± 6 V at 220 K as descriped below.

Transport measurements

The resistance measurements of a ZrNCl-EDLT under both zero and finite magnetic fields were made using a standard four-probe geometry in a Quantum Design Physical Property Measurement System (PPMS) with the Horizontal Rotator Probe with an error below 0.01°, combined with two kinds of AC lock-in amplifiers (Stanford Research Systems Model SR830 DSP lock-in amplifier and Signal Recovery Model 5210 lock-in amplifier). The gate voltage was supplied by a Keithley 2400 source meter. We applied gate voltages to the device at 220 K under high vacuum (less than 10$^{-4}$ Torr), and cooled down to low temperatures. We measured the temperature dependent resistance under magnetic fields (shown in Fig. 2, A and B) under a helium atmosphere (5 ~ 10 Torr) with a cooling speed of 0.1 K/min. All measurements have been performed under the condition that the source-drain current, $I_{DS}$, was less than 0.2 $\mu$A, at which we confirmed that the behavior of $R$ ($T$, $H$, $\theta$) was completely unchanged, compared with that at much smaller $I_{DS}$ (10 ~ 20 nA).

**Supplementary Text**

1. Possibility of electrochemical process

In our whole measurement, the electrochemical process (reaction or intercalation) can be excluded for the following reasons.

First of all, the device was covered with a resist (for electron beam lithography) except for the channel area. This excludes the possibility of intercalations from the edges of ZrNCl crystals. We use a thin flake without any monolayer steps on the surface, and the intercalation from the top surface thus is also forbidden (*14*).

Second, at low temperatures, the activation energy of electrochemical process is significantly suppressed. Figure S1 displays the temperature dependent source-drain and gate current in an EDLT device. The gate current shows an activation-type reduction with temperature, with the activation energy of ~ 0.4 eV. This dramatic reduction of the gate current means the broadening of chemical window (*44*). By using a ZnO-EDLT, it is reported that the maximum gate voltage applicable increases from 3 to 5.5 V, simply by decreasing temperature from 300 K to 220 K (*45*). This also indicates that the effective electrochemical window is broadened by simply reducing temperature. Furthermore, to confirm the reversibility of ionic-liquid gating, we measured transfer curves of ZrNCl-EDLTs at 220 K. Figure S2 shows the source-drain and gate current as a function of gate voltage between 0 and 6 V. The source-drain and gate current of both devices completely returned back to the original values, although the threshold voltages are different for devices.

Finally, as shown in Fig. S3, atomically flat surface morphology did not change at all before and after measurements, which clearly indicates that there is no sign of electrochemical reaction or degradation by applying gate voltages up to 4 – 6.5 V at 220 K.

2. Analyses of the superconducting fluctuation

The analyses of $R_{\text{sheet}}(T)$ curves above $T_c$ were performed based on the model of superconducting amplitude fluctuation (*46*), using the expression,

$$R_{\text{sheet}}(T) = \left( \frac{1}{R_{\text{N}}(T)} + \Delta G_{\text{SF}} \right)^{-1}, \quad \text{(S1)}$$

where $R_{\text{N}}(T)$ is the normal state sheet resistance and $\Delta G_{\text{SF}}$ is the excess sheet conductance due to the superconducting fluctuation (SF). In this work we adopted the formula,

$$\frac{1}{R_{\rm N}(T)} = \frac{1}{a+bT^2} + c\ln\left(\frac{T}{T_0}\right), \tag{S2}$$

where $a$, $b$, $c$ and $T_0$ are $T$-independent fitting parameters. The second term in Eq. S2 comes from the contributions of the weak localization (WL) and the Coulomb interaction between particles with nearly the same momenta (ID), which are remarkable in a dirty 2D system (*47*). For $V_{\rm G}$ = 6.5 V, we obtained the function form of $R_{\rm N}(T)$ by fitting the $R_{\rm sheet}(T)$ curve at $\mu_0 H^\perp$ = 9 T, where the contribution of SF can be ignored since $\mu_0 H_{c2}^\perp(0) \sim 2$ T, with $a = 1.92\times10^2$ $\Omega$, $b = 8.25\times10^{-3}$ $\Omega/K^2$, $c = 3.47\times10^{-4}$ $\Omega^{-1}$, and $T_0 = 52.8$ K. We also checked that the first Drude term in Eq. S2 can reproduce the $R_{\rm sheet}(T)$ curve at $\mu_0 H^\perp = 0$ T above $T = 50$ K, where the quantum contributions (logarithmic term due to WL & ID) are negligible.

According to (*46*), SFs commonly consists of three principal contributions: the Aslamazov–Larkin (AL) process, corresponding to the direct effect of the thermally fluctuating Cooper pairs (*32*), the anomalous Maki–Thompson (MT) process, which is described as an indirect effect of the Cooper pairs after decaying into pairs of quasiparticles (*33*, *34*), and the depression in the electronic density of states (DOS) due to their involvement in fluctuation pairings (*48*, *49*). Using these contributions, the conductance due to SFs is given by (*46*)

$$\Delta G_{\rm SF} = \Delta G_{\rm AL} + \Delta G_{\rm MT} + \Delta G_{\rm DOS} = \frac{e^2}{16\hbar}\left(\frac{T_{c0}}{T-T_{c0}}\right) + \frac{e^2}{8\hbar}\frac{T_{c0}}{T(1-\delta)-T_{c0}}\ln\left(\frac{T-T_{c0}}{\delta T}\right), \tag{S3}$$

where $T_{c0}$ is the temperature at which the finite amplitude of the order parameter develops, and $\delta$ is the pair-breaking parameter. As shown in Fig. S4, the experimental data is well fitted by Eq. S1, taking Eq. S2 and Eq. S3 into account, when $\delta = 0.05$ and $T_{c0} = 14.65$ K. We note that the value of $\delta = 0.05$ is relatively larger when considering the small normal sheet resistance of $\sim 200$ $\Omega$ (*45*). Its origin could be the strong spin orbit interactions generated by the broken inversion symmetry due to the external electric field.

### 3. Temperature dependence of the magnetoresistance

Figure S5 shows the temperature dependence of the magnetoresistance in the perpendicular and parallel magnetic field geometry. In contrast to the parallel magnetic fields, which do not change the zero resistance state at a low temperatures until 9 T, the effect of perpendicular magnetic fields is dramatic, destroying superconductivity at very weak magnetic fields. The drastic increase in the slope of $R_{sheet}(H)$ near $H = 0$ with increasing temperature is also the typical behavior of a 2D superconductor under perpendicular fields. Similar observation has been reported in the previous study (*14*) and other electric-field-induced superconductors (*3*, *15*, *17*). At the same time, we also note that the increasing rate of $R_{sheet}(H)$ is the sharpest around the midpoint, indicative of the validity of our definition of $H_{c2}$. In the in-plane magnetic field geometry, on the other hand, $R_{sheet}(H)$ shows a broad resistive transition only in the limited temperature range around $T_c(0)$. This is the natural consequence of the tiny shift of $R_{sheet}(T)$ by applying the magnetic field in this geometry as shown in Fig. 2B.

4. Hall-effect measurements

We performed Hall-effect measurements under magnetic fields, $\mu_0 H$, of up to 9 T at 60 K to confirm the sign of the charge carrier and the sheet carrier density, $n_{2D}$. Figure S6 shows the anti-symmetrized Hall resistance, $R_{xy}$, of a ZrNCl-EDLT, plotted as a function of magnetic field. $R_{xy}$ is always negative, indicating electron-type carriers, which is consistent with the gate voltage dependence of the sheet resistance (Fig. 1B). $R_{xy}$ clearly shows the $H$-linear dependence up to 9 T at each $V_G$ (black dashed lines). Therefore, we directly extracted $n_{2D}$ from the Hall coefficient $R_H$ for each gate voltage by using the relation $R_H = \dfrac{1}{n_{2D} e}$.

3. Confirmation of the surface morphology after EDLT measurements

We took optical microscope images (Fig. S3, A and B) of one typical ZrNCl-EDLT device before and after the transport measurements including the gate voltage scan as well as atomic-force microscope (AFM) images (Fig. S3, C–F). Under ionic liquid, the device was covered by a resist (ZEP 520 A) except for the channel area (Fig. S3A). After the EDLT experiments, we removed the resist (Fig. 3B) and took some AFM images. Some unclear lines in Fig. 3C are derived from the noise generated by the AFM. Figure S3D is an expanded view of the area enclosed by the yellow rectangle in Fig. S3C. Figure S3E shows a cross-sectional view along the blue line in Fig. S3D. Figure S3F shows a three-dimensional plot of Fig. S3D. From these AFM images, we find that the root-mean-square of the surface roughness is ~ 0.068 nm, which is much smaller than the superconducting thickness $d_{sc} \cong 1.8$ nm, and the thickness of one unit cell of ZrNCl (~ 2.7 nm). These results exclude possible inhomogeneous superconducting states caused by surface roughness, such as granular superconductivity where isolated superconducting islands are weakly coupled. This proves that the ZrNCl-EDLT has little disorder, suggesting weak pinning in this system.

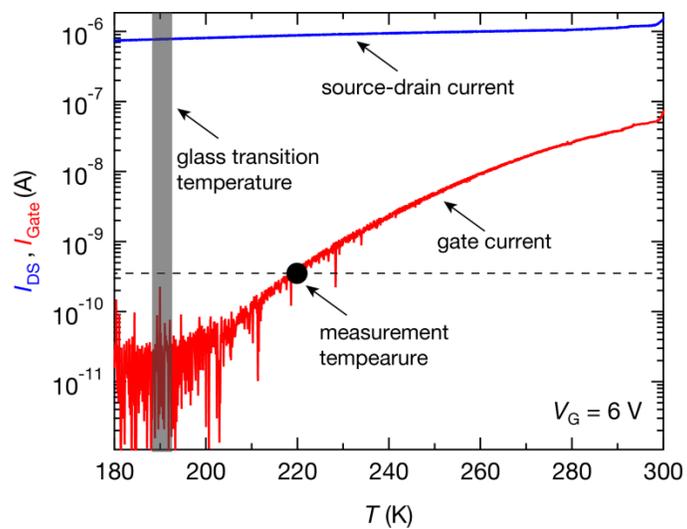

**Fig. S1. Temperature dependence of source-drain current ($I_{DS}$) and gate current ($I_{Gate}$).**

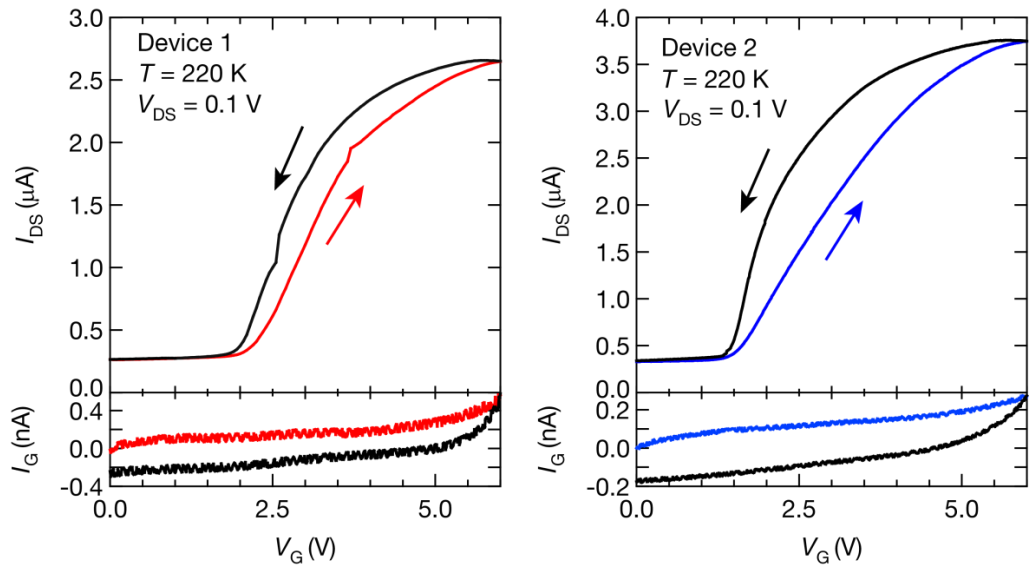

**Fig. S2.** Source-drain current $I_{DS}$ and gate current $I_G$ as a function of gate voltage $V_G$ at 220 K in two different ZrNCl-EDLTs.

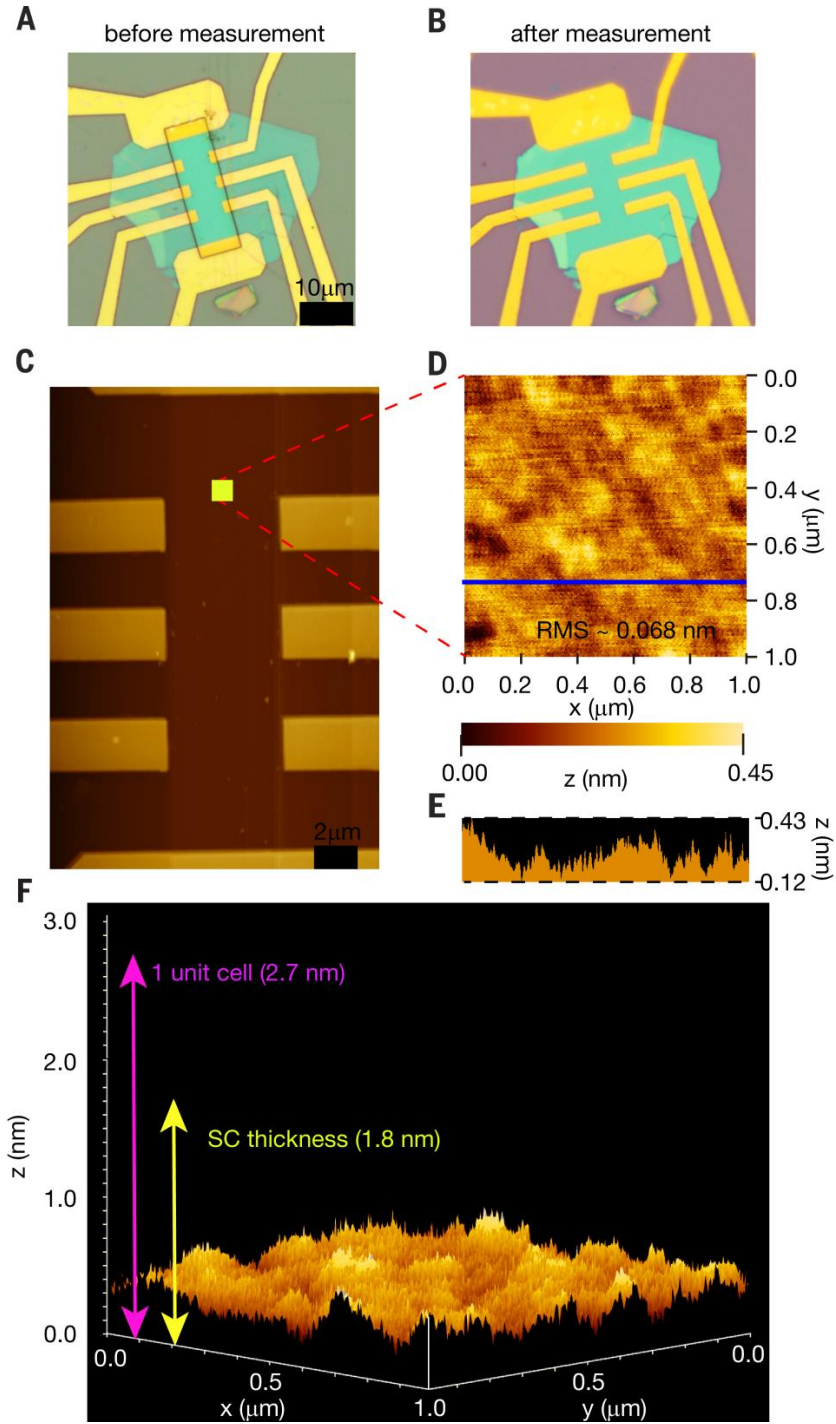

**Fig. S3. Surface morphology of a ZrNCl device after EDLT measurements.** Optical microscope images of the ZrNCl device (**A**) before and (**B**) after transport measurements. (**C**) AFM images around channel area, and (**D**) an expanded view of the area enclosed by the yellow rectangle in Fig. S3C. (**E**) A cross-sectional view through the blue line in Fig. S3D. (**F**) Three-dimensional plot of Fig. S3D.

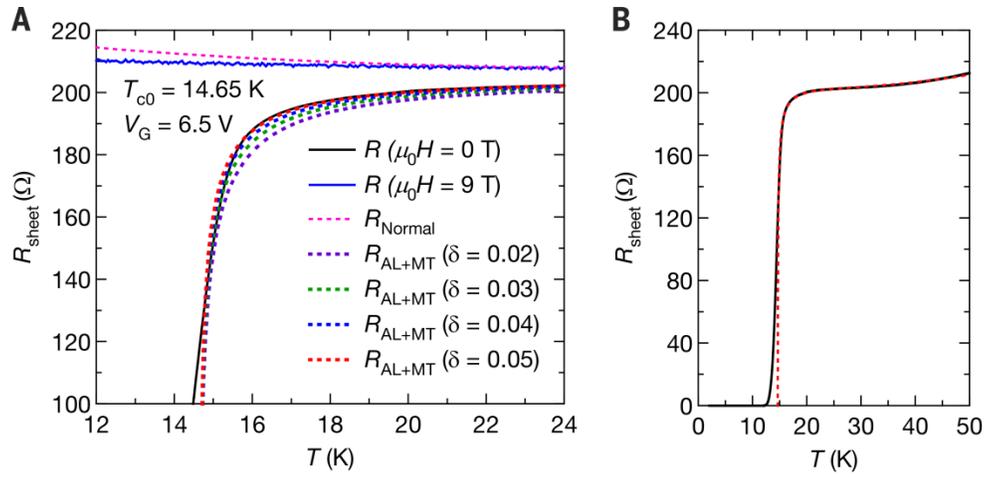

**Fig. S4. Analyses of the superconducting fluctuation.** Temperature dependence of the sheet resistance, $R_{sheet}$, plotted in the ranges (**A**) $T = 12$ to 24 K and (**B**) 0 to 50 K. The solid and dashed lines show experimental data and fitting results, respectively.

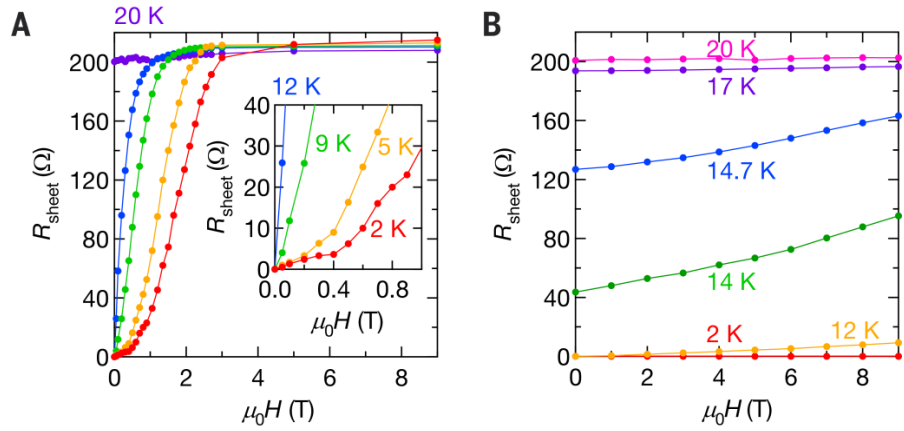

**Fig. S5. Temperature dependence of the magnetoresistance in (A) the perpendicular and (B) the parallel magnetic field geometry.** The inset of Fig. S5A show a close-up of the region near the zero magnetic field and zero resistance.

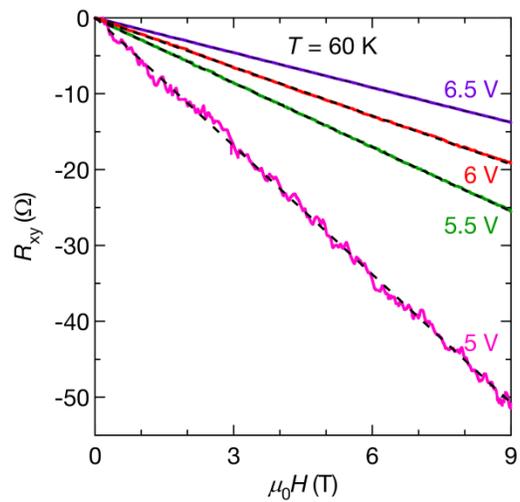

**Fig. S6. Hall-effect measurements on a ZrNCl-EDLT.** Hall resistance, $R_{xy}$, as a function of magnetic field at various gate voltages at 60 K.